\documentstyle[aps,twocolumn,epsf]{revtex}

\begin{document}

\draft

\title{
Quantum Communication with Phantom Photons
}

\author{S.J. van Enk$^1$, H.J. Kimble$^1$, J.I. Cirac$^2$, and P. Zoller$^2$}

\address{$^1$Norman Bridge Laboratory of Physics, California Institute of Technology 12-33,
Pasadena, CA 91125}
\address{$^2$Institut f\"ur Theoretische Physik, Universit\"at Innsbruck,
Technikerstrasse 25, A-6020 Innsbruck, Austria}

\date{\today}

\maketitle
\begin{abstract}
We show that quantum information may be transferred between atoms in different locations by using ``phantom photons'': the atoms are coupled through electromagnetic fields, but the corresponding field modes do not have to be fully populated.   
In the case where atoms are placed inside optical cavities, errors in quantum information processing due to photon absorption inside the cavity are diminished in this way.  
This effect persists up to intercavity distances of about a meter for the current levels of cavity losses, and may be useful for distributed quantum computing.
\end{abstract}
\narrowtext
\section{Introduction}
A standard scheme to transfer population in an atom or molecule from one ground state to another through an excited state employs a sequence of laser pulses that first connects the {\em final} state with the intermediate one, and only then the initial state \cite{counter}. 
If this process is executed adiabatically, the atom stays in a (field-dependent) superposition of ground states, and
ideally the intermediate state is never populated, thus strongly suppressing decays out of the usually unstable excited state, while still achieving population transfer with almost 100\% efficiency.
In Quantum Communication one attempts to transfer quantum information from one location to another over a usually noisy channel. If one could similarly avoid populating the intermediate noisy state, this would be useful both for distributed quantum computing and for quantum cryptography, as losses due to photon absorption could be partially avoided this way, and as an eavesdropper would not be able to use eavesdropping techniques that rely on the actual presence of the photon. Although the two situations are different, the mathematical descriptions are sufficiently similar to suggest that the effect should exist in quantum communication as well. Here we investigate this idea in a realistic physical setup.

In \cite{Cirac} a physical implementation for communication in a quantum network has been proposed, based on high-Q cavity QED: atoms inside cavities store quantum information while photons, produced by laser manipulation of the atoms, are the data buses, carrying information from one cavity to another. 
In that proposal, laser pulses were designed in such a way that one atom-cavity systemproduces a time-symmetric photon wave packet, so that the seemingly unavoidable 
reflection of the photon wave packet from the almost perfectly reflecting mirror of a second distant cavity is prevented.
In the same setup, a different idea \cite{Pelli} is to use adiabatic passage to accomplish the coherent transfer of quantum information through a dark cavity state: ideally there is never a photon inside the cavity. It is the latter scheme that forms the inspiration for the present work: is it possible to communicate between two cavities using a dark ``fiber'' state and thus diminish losses inside the fiber that connects the cavities? Or, stated somewhat differently, can one do without a photon altogether?
Although error correction schemes have been designed specifically for photon absorption errors \cite{Enk1,Enk2,Briegel}, it would be even better to avoid the error.

\section{A physical implementation for quantum communication}

The physical system under consideration consists of two high-Q optical cavities connected by a quantum channel (which can be an optical fiber for longer distances, or just the vacuum for shorter distances), which will be called the ``fiber'' from now on. We choose the optical frequency domain because thermal optical photons are practically absent at room temperature.
There are (at least) two ways of describing the electromagnetic field modes in the presence of the cavities and fiber. One description defines cavity modes and fiber modes separately, each mode being represented by creation and annihilation operators that satisfy the standard commutation relations. There is a
linear coupling arising from the fact that the cavity mirrors are not perfect and transmit a small fraction of the incident light from cavity to fiber and {\em vice versa}. 
This description has an obvious interpretation, but the definition of cavity and fiber modes is valid in principle only for high-finesse systems, for time scales long compared to the roundtrip time, and for operation close to resonance (for instance, only then standard commutation rules are satisfied \cite{Vogel}). 
Therefore, although we will use the terminology corresponding to the picture of cavity and fiber modes, in the calculations we will instead employ a different, rigorous description  in terms of second-quantized fields in the complete system consisting of 2 cavities placed at a distance $L$. One goal of this paper is to extend the analysis of \cite{Pelli}, which used the simpler description.
  
For each longitudinal field mode there are in principle infinitely many transverse modes. However, the spherical mirrors of the cavity lift the degeneracy of those transverse 
modes by such a large amount that only one mode is nonnegligibly coupled to an atom inside the cavity. This single transverse mode in turn couples only to a single transverse mode outside the cavity \cite{note1}. This situation justifies, for our purposes, using a 1-dimensional model where each mode is characterized by a single wave number $k$.

We thus model the two cavities of length $l$ at distance $L$ by two perfect mirrors located at $z=\pm(L/2+l)$ and two highly reflecting (but not perfect) mirrors located at $z=\pm L/2$. The transmission coefficients of these imperfect mirrors are taken to be identical and equal to
$t(k)=2/(2-i\mu k)$ with $\mu$ real, and the reflectivity is then $r(k)=1-t(k)$. 
This model for a mirror corresponds to that of a dielectric with thickness $d$ and dielectric constant $\epsilon$ in the limit $d\rightarrow 0$ and $\mu=d\epsilon$ constant. A photon inside such a cavity leaks out through the partially transparent mirror at a rate $\kappa_c=c|t|^2/(2l)$.
\subsection{Field modes}
The normal modes $U_k(z)$ and $U'_k(z)$ for a system consisting of just the two imperfect mirrors located at $z=\pm L/2$ have been calculated in \cite{Loudon} (Eqs. (11) and (12)). 
We need certain combinations of these modes, denoted by $V(z)$, that satisfy 
additional boundary conditions $V(z)=0$ at $z=\pm(L/2+l)$. There are two types of such modes, which we refer to as even and odd modes resp., which can be written as  
\begin{equation}
V_k^{\pm}(z)=N_{k,\pm}^{-1/2}\left[U_k(z)\mp U'_k(z)\right],
\end{equation}
for each $k$ that is a solution of
\begin{equation}
\exp(-2ikZ)=\pm T_k-R_k,
\label{Eq}\end{equation}
where $Z=L/2+l$ and $R_k$ and $T_k$ are reflection and transmission 
coefficients as defined in Eqs. (15--19) of \cite{Loudon}. 
Inside the two cavities, the mode functions take the simple form
\begin{eqnarray}\label{inside}
U_k(z)\mp U'_k(z)&=&2 \sin [k(z+Z)]\,\,\,{\rm for}\,\,\, z<-Z+l,\nonumber\\
U_k(z)\mp U'_k(z)&=&\mp 2 \sin [k(z-Z)]\,\,\,{\rm for} \,\,\, z>Z-l,
\end{eqnarray}
so that the boundary conditions are indeed fulfilled.
The normalization factors are 
\begin{eqnarray}
N_{k,\pm}&=&\int_{-Z}^Z |U_k(z)\mp U'_k(z)|^2 {\rm d} z\nonumber\\
&\equiv&2N^c_{k,\pm}+N^f_{k,\pm},
\end{eqnarray}
where $N^{c,f}_{k,\pm}$ give the cavity and fiber contributions,
\begin{eqnarray}\label{norms}
N^c_{k,\pm}&\equiv&\int_{-Z}^{-Z+l} |U_k(z)\mp U'_k(z)|^2 {\rm d} z=2l-\frac{\sin(2kl)}{k}\nonumber\\
N^f_{k,\pm}&\equiv&\int_{-Z+l}^{Z-l} |U_k(z)\mp U'_k(z)|^2 {\rm d} z=\nonumber\\
&&\left[L\mp\frac{\sin(kL)}{k}\right]\frac{|t|^2}{|1\pm r\exp(ikL)|^2}.
\end{eqnarray}
Eq.~(\ref{Eq}) has been solved numerically, and typical results for the case we are interested in, with $l\ll L$ and $|t|\ll 1$, are displayed in Fig.~1. 
From now on we will for convenience denote the modes by the index $i\equiv k,\pm$.
\subsection{Interaction with atoms}
Inside each of the 2 cavities we have one atom, denoted by $A$ and $B$ respectively, that we use to store qubits. Both atoms are assumed to possess 2 ground states, denoted by $|0\rangle$ and $|1\rangle$. The field modes $i$ couple state $|0\rangle$ to an excited state $|e\rangle$ with coupling strengths $g_i$, which depend on the positions of the atoms:
For an atom at position $z=\pm(Z-s)$ for $0< s <l$ (i.e. inside one of the two cavities) we have
\begin{equation}
g_{i}(s)=\sqrt{\frac{1}{N_i}}\times\sqrt{
\frac{d^2\omega_0}{2\hbar\epsilon_0{\cal A}}}\times(\mp 1)^{n}\sin(ks),
\label{gi}\end{equation}
where ${\cal A}$ is the area of the light beam and $d$ and $\omega_0$ are
the dipole moment and resonance frequency of the atom. 
The phase factor $(\mp 1)^{n}$ with $n$ even for the even modes and odd for the odd modes arises from the corresponding phase factor in the mode functions inside the two cavities (cf. Eq. (\ref{inside})). 
As expected, an atom inside one of the cavities couples mainly to the modes with a large ``cavity mode'' content, for which $N^f_{i}$ and thereby $N_{i}$ becomes small. 
The state
$|e\rangle$ in turn is coupled to state $|1\rangle$ by a laser field of different polarization at frequency $\omega_L$ with a Rabi frequency $\Omega(t)$. 
In order to diminish the effects of spontaneous emission from the excited state $|e\rangle$
(at a rate $\gamma_e$) the laser detuning from that state is taken to be much larger than all other rates in the problem. In particular we take $\Delta\equiv\omega_L-\omega_0\gg\gamma_e$, and henceforth we neglect $\gamma_e$.

The same condition on $\Delta$ justifies eliminating the upper state adiabatically, and the Hamiltonian describing the interaction between the two ground states and the field modes is
 then (in a frame rotating at the laser frequency)
\begin{eqnarray}
H=\sum_i\frac{g_i^2}{\Delta}
a_i^{\dagger}a_i |0\rangle\langle 0|
+\delta\omega(t)|1\rangle\langle 1|+ \nonumber\\
\sum_iG_i(t)\left[e^{i\phi(t)}|1\rangle\langle 0| a_i+{\rm h.c.}\right],
\label{H}\end{eqnarray}
where the summation is over all modes $i$,  $\delta\omega(t)=\Omega(t)^2/(4\Delta)$ is the AC-Stark shift due to the laser field, $g_i$ is the coupling constant with mode $i$, as given in (\ref{gi}),
and the effective coupling between the two ground states through mode $i$ is $G_i(t)=g_i\Omega(t)/(2\Delta)$.
As in \cite{Cirac}, the laser phase $\phi(t)$ can be chosen to compensate for the time-dependent shift $\delta\omega(t)$ of $|1\rangle$.

The total Hamiltonian then consists of the sum of the free (effective) Hamiltonian for the field modes, $H_{{\rm field}}=\sum_i -\delta_i a_i^{\dagger}a_i$, where $\delta_i=\omega_L-ck_i$ 
is the detuning from mode $i$, and terms like (\ref{H}) for atoms $A$ and $B$.
The quantities $k_i$ are determined by numerical solution of (\ref{Eq}). The easiest way to describe the evolution of our system is in the Schr\"odinger picture, as in the case of our interest the number of excitations is always at most 1. 
Thus we can conveniently denote by $|1\rangle_i$ the state where all field modes are empty except for mode $i$ which has one photon, and by $|0\rangle_i$ the vacuum state of the field. 
The initial  state $|\psi_{100}\rangle=|1\rangle_A|0\rangle_i|0\rangle_B$ is coupled to a set of intermediate states where one photon is in one of the modes $i$, $|\psi^i_{010}\rangle=|0\rangle_A|1\rangle_i|0\rangle_B$, which in turn are coupled to the desired final state, where the excitation has been transferred to atom $B$: $|\psi_{001}\rangle=|0\rangle_A|0\rangle_i|1\rangle_B$.
The evolution equations in the interaction picture for the corresponding amplitudes $c_{100}$, $c^i_{010}$ and $c_{001}$ are given by
\begin{eqnarray}
i\dot{c}_{100}&=&\sum_i G^A_i(t)c^i_{010},\nonumber\\
i\dot{c}_{001}&=&\sum_i G^B_i(t)c^i_{010},\nonumber\\
i\dot{c}^i_{010}&=&-\delta_ic_{010}^i+G^A_i(t)c_{100}+G^B_i(t)c_{001},
\label{set}\end{eqnarray}
where the index $A,B$ refers to the atoms $A$ and $B$.
The AC-Stark shifts due to the atomic coupling $g_i$ to the modes $i$ have been absorbed into the amplitudes $c^i_{010}$, and similarly the phases $\phi_{A,B}$ have been absorbed into the amplitudes $c_{100}$ and $c_{001}$, respectively. Note that there is no delay time $L/c$ appearing explicitly in these equations; nevertheless, since the whole mode structure derived from Maxwell's equations is present,  this delay is incorporated implicitly, and hence causality will not be violated.

Finally, let us return to the phase factor $(\mp 1)^{n}$ in the coupling coefficients (\ref{gi}). 
The alternating sign of the coupling is not harmless. In fact, it prevents the existence of a perfect dark state in this multimode system: a superposition of ground states not coupled to the even modes, is coupled to the odd modes, and {\em vice versa}.
We can compare this situation with that of population transfer in an atom: we recall that population transfer through a {\em single} continuum of intermediate states is very well possible (in fact, in a simple model case the transfer is found to be complete \cite{hioe}, but in reality there inevitably are effects which will reduce the benefits of adiabatic passage through a continuum \cite{takashi}). However, the situation at hand is, at least for large intercavity distances, like that of an atom with {\em two} independent continua (corresponding to the odd and even modes, resp.), where any beneficial interference effects are canceled out.
Thus for large $L$ (a condition to be made more precise below), where the mode structure approaches that of a continuum,  we do not expect to be able to avoid losses to any degree, irrespective of whether the losses are inside the fiber or the cavity.
\subsection{Introduction of losses}
As losses, especially due to photon absorption, are inevitable, the important question will be what the influence of losses is on the transfer of quantum information. 
There are essentially two types of losses: {\em (i)}
Photons are irreversibly lost inside the cavity (this includes photons leaking out of the other side of the cavity) at a rate $\gamma_c$. Thus, a fraction $\gamma_c/(\gamma_c+\kappa_c)$ of the photons is lost in each of the two cavities (the remaining fraction $F_c\equiv\kappa_c/(\kappa_c+\gamma_c)$ going from cavity to fiber). 
{\em (ii)} Photons inside the fiber are absorbed at a rate $\gamma_f=\alpha c$ in terms of the absorption coefficient $\alpha$ of the fiber.
The fraction of photons surviving travel through a fiber of length $L$ is $\exp(-\alpha L)$. 

In terms of the modes $i$, which are mixed cavity/fiber modes, the losses are taken into account by a decay rate $\gamma_i$ of each mode, which is just a weighted average of the fiber and cavity loss rates,
\begin{equation}\label{weight}
\gamma_i=2N^c_i\gamma_c+N^f_i\gamma_f,
\end{equation}
where the normalization factors were defined in (\ref{norms}). This relation can be derived from modeling the losses as due to a large set of absorbers (atoms) inside cavities and fiber off-resonantly coupled to the modes $i$.
Namely, in that case each field mode 
effectively couples to a continuum, which introduces both a decay
 $\gamma_i$ and an energy shift $S_i$ according to
\begin{equation}
S_i-i\gamma_i/2\equiv\sum_j\int_{-Z}^{Z}\frac{|\mu_j|^2}{\Delta_{ji}+i\Gamma_j/2}|V_i(z)|^2 \rho_j(z){\rm d}z,
\label{Ci}\end{equation}
where the sum is over different types of atoms or atomic levels $j$ involved, $\mu_j$ is the coupling constant to level $j$, $\Delta_{ji}$ the detuning of mode $i$ from level $j$, which we approximate to be independent of $i$, $\Gamma_j$ the decay rate of level $j$, while $\rho_j(z)$ gives the number density of atoms with levels $j$ at position $z$. The result (\ref{weight}) follows from (\ref{Ci}) if we assume the densities $\rho_j(z)$ to be piecewise constant inside the fiber and inside the two cavities.
We may neglect the shifts $S_i$ if we assume $\Delta_j$ to take both positive and negative values. In addition to the decay rates, cross couplings between different modes $i$ and $i'$ exist as well, given by 
\begin{equation}\label{Cii}
{\cal C}_{ii'}\equiv\sum_j\int_{-Z}^{Z}\frac{|\mu_j|^2}{\Delta_j+i\Gamma_j/2}V_i(z)V^*_{i'}(z) \rho_j(z){\rm d}z.
\end{equation}
We may again neglect the real part of ${\cal C}_{ii'}$, but the imaginary part is nonzero, and can in fact be related to $\gamma_f$ and $\gamma_c$: using that $l\ll L$ we have 
\begin{equation}
{\cal C}_{ii'}=i(\gamma_f-\gamma_c)\sqrt{\frac{2N^c_i}{N_i}\frac{2N^c_{i'}}{N_{i'}}}
\,(i\neq i')
\end{equation}
for modes with the same parity, and ${\cal C}_{ii'}=0$ for modes of opposite parity.
Here we used the orthogonality of the modes $i$: for instance,
for $\gamma_c=\gamma_f$ we have $C_{ii'}=0$ for all modes, as follows directly from (\ref{Cii}) and the orthogonality relations. 

We note that with the assumption of a constant absorber density in the fiber, the number of atoms involved in fiber decay increases linearly with the fiber length $L$. Since at the same time the coupling
to each atom goes down with $\sqrt{L}$ the decay rate of each ``fiber'' mode is indeed
independent of $L$, as we already tacitly assumed.  
For simplicity, we take decay due to losses through the outside mirrors at $z=\pm (L/2+l)$ into account by the same equation (\ref{weight}) for $\gamma_i$, as the respective normalization factors can be interpreted as the probabilities of finding a photon from mode $i$ in the fiber or inside the cavity, and since it is only the cavity mode that decays through the outside mirrors.

In the presence of losses, the equations (\ref{set}) for $c_{100}$ and $c_{001}$ are still valid, but the third equation is replaced by
\begin{eqnarray} \label{loss}
i\dot{c}^i_{010}&=&-(\delta_i+i\gamma_i)c_{010}^i+\sum_{j\neq i} {\cal C}_{ij} c_{010}^j+\nonumber\\&&
G^A_i(t)c_{100}+G^B_i(t)c_{001}.
\end{eqnarray}

\section{Numerical results}

We numerically solve the equations (\ref{set}) and (\ref{loss}). The number of modes $i$ to be included depends on the fiber length $L$. For each calculation we have verified that our results have converged with respect to the number of modes included.

We assume that the laser pulses $\Omega_{A,B}(t)$ are Gaussian in shape, and that the laser pulse in the first cavity is effectively delayed by a time $\tau$ compared to the second pulse, i.e. in real time we have $\Omega_A(t)=\Omega_B(t-L/c+\tau)$.
Thus for $L<\tau/c$ the second pulse is actually turned on first.
The optimum delay time turns out to be $\tau=1.2 T$, where $T$ is the width of the pulses (which here is defined by $\Omega(t)=\Omega_0\exp(-t^2/T^2)$). For optimum transfer $T$ has to be larger than $\kappa_c^{-1}$ and much larger than the inverse Rabi frequencies so as to satisfy adiabaticity requirements: we took $T=20\kappa_c^{-1}$ (except for larger $L$ ($L>l/|t|^2$) where we chose $T=40\kappa_c^{-1}$).
Given $T$, we then optimize the strength of the laser coupling and laser detuning to find the best possible probability 
\begin{equation}\label{P}
P=|c_{001}(t\rightarrow\infty)|^2
\end{equation}
 to transfer a qubit from atom $A$ to atom $B$.

The parameters chosen in all subsequent analysis correspond to present-day technology \cite{Rempe,Hood}. We take a wavelength around  $\lambda=852$nm corresponding to the D2 line in Cs. We take $\mu k_0=500$ for $k_0=2\pi/\lambda$, which implies  $|t|^2=1.6\times 10^{-5}$; we take a cavity of size $l=10^{-5}$m, so that
the cavity decay rate equals $\kappa_c/2\pi=38$ MHz.
The (maximum) coupling between atom and the pure ``cavity mode'' is taken to be $g_{{\rm max}}/2\pi=100$ MHz, and the detuning is $\Delta/2\pi=500$ MHz. 

We first examine to which extent one needs to populate the intermediate states containing one photon to transmit one qubit in the lossless case (i.e. $\gamma_c=\gamma_f\equiv 0$). 
As is typical for the adiabatic passage scheme, the total population in the intermediate states can easily be less than $5\%$ at any time during the entire transmission. The relevant quantity, however, is not the maximum population, which can be made smaller by simply increasing $T$. Instead one has to consider also the total time that that population is present. We therefore consider the following.
In the picture that a photon wave packet is produced inside a cavity and subsequently travels through the fiber, we expect the process to take a total time of $L/c+2\kappa_c^{-1}$:
the time spent by one photon inside the fiber is $L/c$, while the time spent inside each cavity is $\kappa_c^{-1}$ on average. Hence, if we integrate the population in fiber/cavity modes over time, and divide by the amount of population transferred from $|0\rangle_B$ to $|1\rangle_B$ in atom $B$ and by $L/c+2\kappa_c^{-1}$, we expect this number,
\begin{equation}\label{ratio}
R\equiv\frac{\int {\rm d}t\sum_i |c_{010}^i(t)|^2}{|c_{001}(t\rightarrow\infty)|^2}\frac{1}{L/c+2\kappa_c^{-1}},
\end{equation}
to be larger than or equal to 1.
Fig.~2 shows, however, that this number may be below unity, and only around a value of $L\approx L_{{\rm eff}}\equiv l/|t|^2\approx 0.6$m, 
does this number increase from near 0 to 1. At this length, the effective number of modes coupled to the cavity becomes larger than 1, and the same threshold will be found in the presence of losses. 
For smaller $L$, the fact that $R$ is smaller than 1 is a manifestation of a dark state and shows that the intermediate states do not have to be fully populated.
We also note that the ratio (\ref{ratio}) is minimized for maximum transfer of information (in fact \ $|c_{001}(t\rightarrow\infty)|^2\approx 1$ in all cases). 

We now turn to the question of how losses affect the transfer of information,  in particular, whether the fact that the intermediate states do not have to be fully populated can be exploited to reduce losses.
The probability of sending one photon from one cavity to the other
is at most equal to the probability $\exp(-\alpha L)$ of sending a photon through the fiber multiplied by the probability $F_c^2$ that the photon leaves the first cavity and enters the second one. That is, we expect $P$, defined in (\ref{P}), to be at most
\begin{equation}
P_{1}=\left(\frac{\kappa_c}{\kappa_c+\gamma_c}\right)^2\exp(-\gamma_f L/c).
\label{predict}\end{equation}
while multiple reflections between the two cavities will reduce this number.
However, as we will show below, it turns out that the losses in transmission of information can be limited to less than indicated by these numbers: in particular, if the cavity losses are dominant that probability can be strictly larger than $F_c^2$. In the more general situation where $\gamma_c\approx\gamma_f$, the probability to transmit a qubit can be strictly larger than $P_1$.
On the other hand, if the fiber losses become dominant, the optimum probability of transmitting a qubit over this lossy channel will not be larger than $\exp(-\gamma_f L/c)$.

The lowest value for the unwanted cavity loss rate was reported in \cite{Rempe} to be $\gamma_c=2\times 10^{-6}c/(2l)=3\times 10^7 $s$^{-1}$. 
The relation $\kappa_c=8\gamma_c$ has the interpretation that 1 out of 9 cavity photons is lost.
At the best wavelength of 1.55 $\mu$ the loss rate in a fused silica optical fiber is only $\gamma_f=1.4\times 10^{4} $s$^{-1}$. Even at the wavelength of interest here, $\lambda=852$ nm, the fiber loss rate $\gamma_f\sim 3\times 10^5 $s$^{-1}$  is still much smaller than the cavity loss rate. 
In this case the relevant question would be whether one can avoid losses inside the cavity.
Indeed, by tuning on resonance with the mode $i$ that is most cavity-like the optimum transfer is clearly better than $P_{1}=F_c^2$ (this corresponds to  using the dark cavity state as in Ref. \cite{Pelli}). Even better, however, is to tune to the next mode of the same parity: that mode is still strongly coupled to the atoms, but has a smaller cavity mode content and hence decays at a lower rate. This is illustrated in Fig.~3.
 
In order to investigate the intermediate regime, where $\gamma_c=\gamma_f$, we show in Fig.~4 that the optimum transfer also in that case is better than $P_{1}$ for not too long fibers: this again is a manifestation of the (imperfect) dark state. For long fibers $L\gg l/|t|^2$ one reaches $P_{1}$ as a limit. 

On the other hand, for the case where the fiber losses are dominant, our numerical results for $\gamma_f \gg \gamma_c$ (not shown here) indicate one cannot improve upon the standard exponential loss inside the fiber, even if there is never more than, say, $5\%$ probability to actually find a photon inside the fiber at any time during the transmission.

\section{Discussion and conclusions}
In conclusion, quantum information can be transferred from one atom to the next by using ``phantom photons'': the intermediate states where the quantum bit is carried by a photon inside a cavity or inside a fiber need not be fully populated. Neither energy conservation nor causality are violated. 
Losses due to photon absorption inside the cavities can be diminished this way.  
This effect can be viewed as resulting from the fact that one can restrict the number of reflections of the 
light field inside the cavity to below the number of reflections expected, $1/|t|^2$.  
On the other hand, our results indicate that losses inside the fiber cannot be overcome.
The main reason for this \cite{Klaus} is as follows: 
In the absence of losses all modes are non degenerate and can in principle be selectively excited. In particular, one can couple the atom to a mode that has only a small amplitude inside the fiber, but has appreciable amplitudes in both cavities. By using those nonlocal modes one can ``skip'' the fiber. In the presence of losses, however, the modes become more and more degenerate with increasing absorption due to the cross coupling between the different modes. Consequently, the atom couples to combinations of degenerate odd and even modes that are increasingly localized, i.e. the modes live in either one of the cavities, but not in both.  In that case it would seem the fiber can no longer be skipped.
For future research we suggest here a possibility that perhaps allows one to escape this conclusion: by tayloring laser pulses such that its Fourier spectrum contains peaks only at desired mode frequencies one might still be able to prevent fiber losses to some extent, although the argument that the modes become more and more degenerate still applies.

The suppression of losses  for present technical capabilities is most efficient on smaller scales, and therefore most useful for distributed quantum computing, as for instance in the setup consisting of micro-fabricated elliptical ion trap arrays proposed in \cite{DeVoe}. Namely, 
losses can be partially avoided until too many modes are coupled to the cavity modes, i.e., until $L\gg L_{{\rm eff}}=l/|t|^2$, as has been confirmed numerically; for the parameters used here, pertaining to a Cs atom inside a high-finesse Fabry-Perot cavity, this corresponds to $L_{{\rm eff}}\approx 0.6$ m. This restriction is determined by the quality of the cavity through $\kappa_c$.
Larger quality factors and correspondingly lower decay rates $\kappa_c$ and $\gamma_c$ have been measured for fused silica microspheres \cite{Dave}, which would increase the length $L_{{\rm eff}}$ by at least one order of magnitude.
Since, moreover, such microspheres at a wavelength of $\lambda=1.55\mu$ would have  the same low loss rate $\gamma_c$ as fibers, they in principle are even better candidates for implementing quantum communication or distributed quantum computing, with or without phantom photons. 
\section*{Acknowledgments}
One of us (S.J.v.E) wishes to thank Klaus M\o lmer and Xiatra Anderson for many stimulating discussions and dinners during the 'Workshop on Quantum Computation and Quantum Optics' in Benasque (Spain). 
We thank Thomas Pellizzari for his inspiring talk at the ITAMP workshop on 'Experimental Realizations of Quantum Logic' in Boston.

This work was funded by DARPA through the QUIC (Quantum Information and Computing) program administered by the US Army Research Office, the National Science Foundation, and the Office of Naval Research; and by the European TMR network, ERB-FMRX-CT96-0087.

\begin{figure}[htbp]
\label{modes}
  \begin{center}
   \leavevmode
     \epsfxsize=7cm  \epsfbox{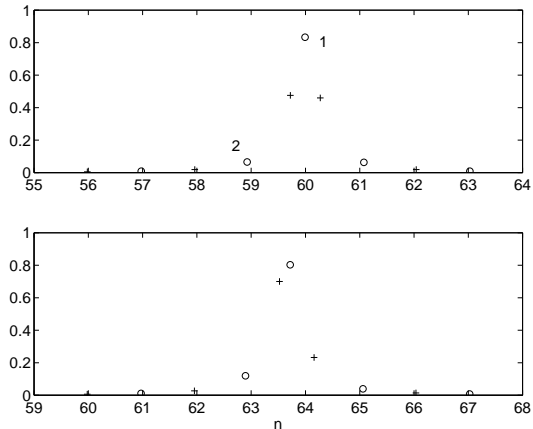}  
\caption{
Cavity mode content $2N^c_{k,\pm}/N_{k,\pm}$ as a function of $n$, which is related to the wave number $k$ by $k\equiv(11\times 10^5 +n)\pi/L$; $n$ is approximately an integer, except around the resonance with the cavity mode (around $n\approx 60$ in this case). Here $L/l=10^5-1/3$ for the upper graph, and $L/l=10^5$ for the lower.
Even (odd) modes are denoted by $+$ (o) and are located at the solutions for $k$ to (\protect{\ref{Eq}}). The even modes of the upper graph indicated by labels '1' and '2' correspond to the two modes used in the calculation for Figure~3. 
}
\end{center}
\end{figure}

\begin{figure}[htbp]
\label{one}
  \begin{center}
   \leavevmode
     \epsfxsize=7cm  \epsfbox{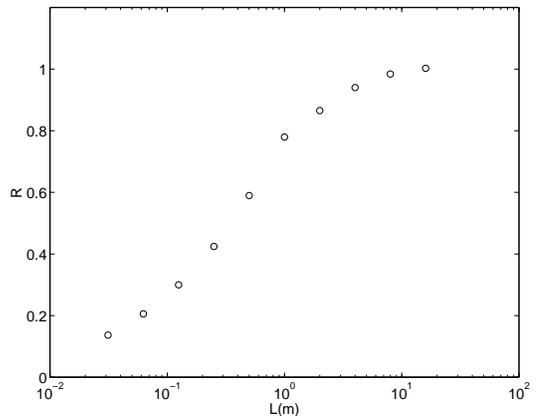}  
\caption{The time spent on average by photons in the fiber and the cavities, normalized by the expected time $2\kappa_c^{-1}+L/c$, as a function of $L$.}
  \end{center}
\end{figure}

\begin{figure}[htbp]
\label{threep}
  \begin{center}
   \leavevmode
     \epsfxsize=7cm  \epsfbox{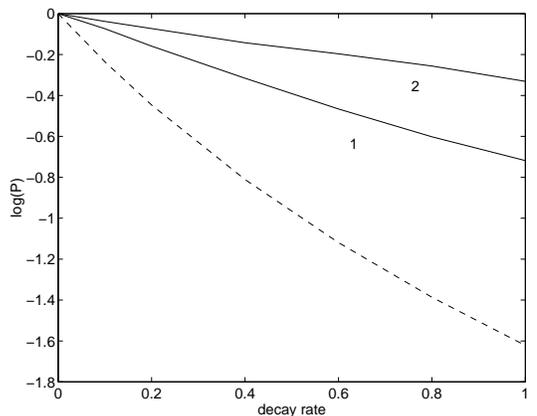}  
\caption{Logarithm of the probability $P$ to transfer a qubit as a function of $\gamma_c$ in units of $c/L_0$ with $L_0=1$m where $\gamma_f=0$. The dashed line gives $\log(P_{1})=2\log(\kappa_c/(\kappa_c+\gamma_c))$ as a reference.  Curve 1 corresponds to tuning the laser on resonance with the most cavity-like mode, curve 2 to tuning the laser around a neighboring mode with the same parity, as indicated in Figure~1. 
}
  \end{center}
\end{figure}

\begin{figure}[htbp]
\label{four}
  \begin{center}
   \leavevmode
     \epsfxsize=7cm  \epsfbox{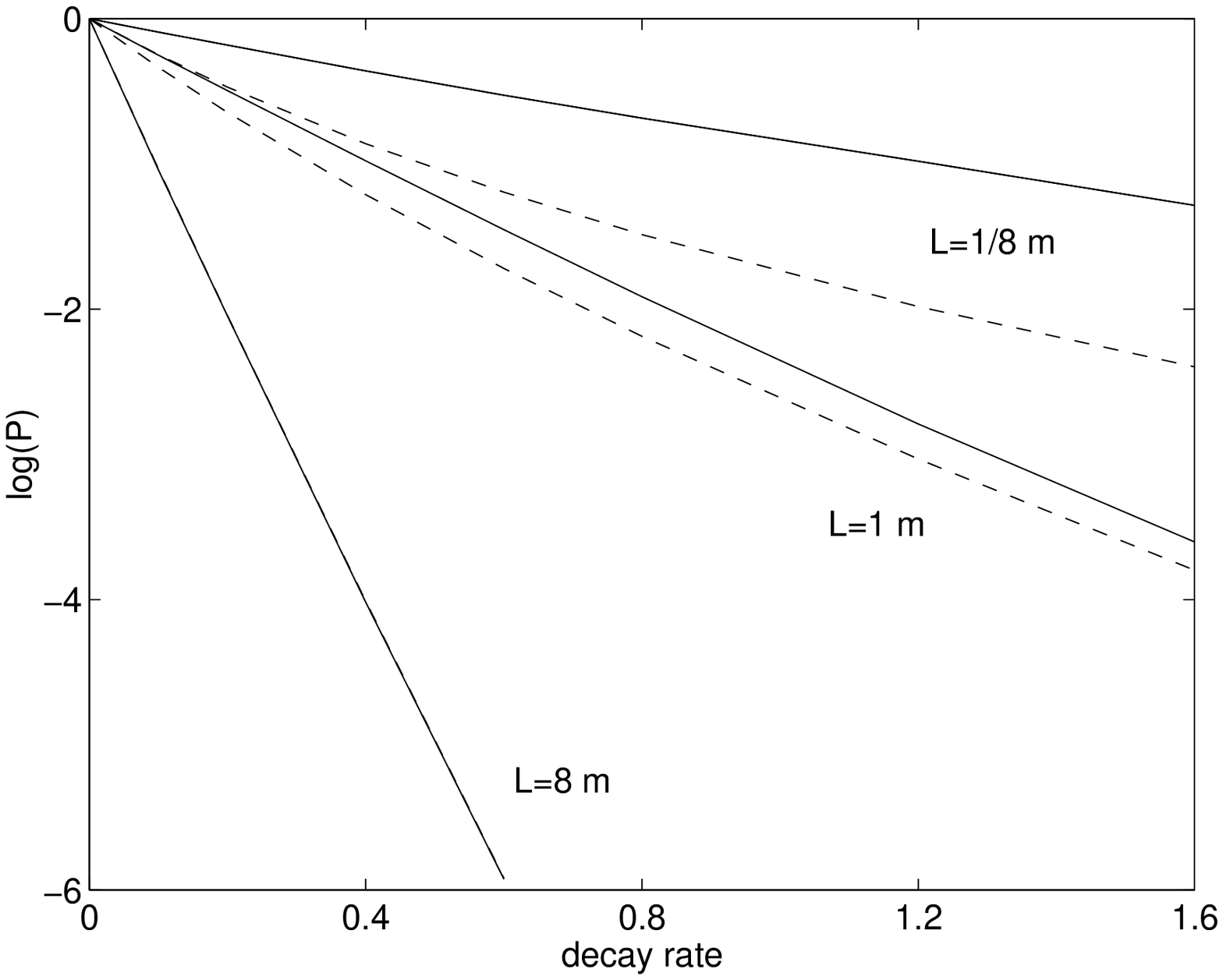}  
\caption{Logarithm of the probability $P$ to transfer a qubit as a function of $\gamma_f$ in units of $c/L_0$ for several values of $L$, for $\gamma_c=\gamma_f$. The dashed lines give $\log(P_{1})$ as references. Note that for $L=8$m the solid and dashed curves fall on top of each other.
}
  \end{center}
\end{figure}

\begin{references}
\bibitem{note1}This mode is just the continuation outside the cavity of the transverse mode inside the cavity. Since this mode will diverge in free space, a series of lenses (or indeed a waveguide such as a fiber) is needed to propagate that mode and couple it to the next cavity.
\end{references}
\end{document}